\documentclass{article}
\usepackage[numbers,sort&compress]{natbib}
\usepackage{spconf,amsmath,graphicx}
\usepackage{amssymb}
\usepackage{amsfonts}
\usepackage{todonotes}
\usepackage{multirow}
\usepackage{bm} 
\usepackage{makecell}

\newcommand{\vf}{\mathbf}

\DeclareMathOperator*{\argmax}{arg\,max}

\title{End-to-end Alexa device arbitration}
\name{Jarred Barber$^{1*}$, Yifeng Fan$^{2*\dagger}$, Tao Zhang$^1$
\thanks{$^*$Equal contribution.}\thanks{$^\dagger$Work performed while at Amazon Alexa AI.}
}
\address{$^1$Amazon Alexa AI, USA \\ $^2$ University of Illinois, Urbana-Champaign, USA}
\begin{document}
\maketitle
\begin{abstract}
    We introduce a variant of the speaker localization problem, which we call \emph{device arbitration}. In the device arbitration problem, a user utters a keyword that is detected by multiple distributed microphone arrays (smart home devices), and we want to determine which device was closest to the user. Rather than solving the full localization problem, we propose an end-to-end machine learning system. This system learns a feature embedding that is computed independently on each device. The embeddings from each device are then aggregated together to produce the final arbitration decision. We use a large-scale room simulation to generate training and evaluation data, and compare our system against a signal-processing baseline.
\end{abstract}
\begin{keywords}
keyword-spotting, speech-recognition, source-localization
\end{keywords}
\section{Introduction}
\label{sec:intro}
A growing number of households own multiple smart speaker voice-assistant devices, such as Amazon Alexa or Google Home. These devices typically operate by listening for a specific wakeword (such as ``Alexa'' or ``Hey Google''; also referred to as a \emph{keyword} or \emph{hotword}) to trigger processing user speech and generating responses. 
When multiple devices are present, they may all detect the same wakeword, creating ambiguity about which device is the correct one to service the user's request. While there may be specific context relevant to the request (for example, a request to play a video should be serviced by a video-capable device), we focus on a heuristic where we want to choose the device that is \emph{physically closest to the user}. We refer to this process as \emph{device arbitration}.
\begin{figure}[b!]
  \centering
  \includegraphics[width=0.7\linewidth]{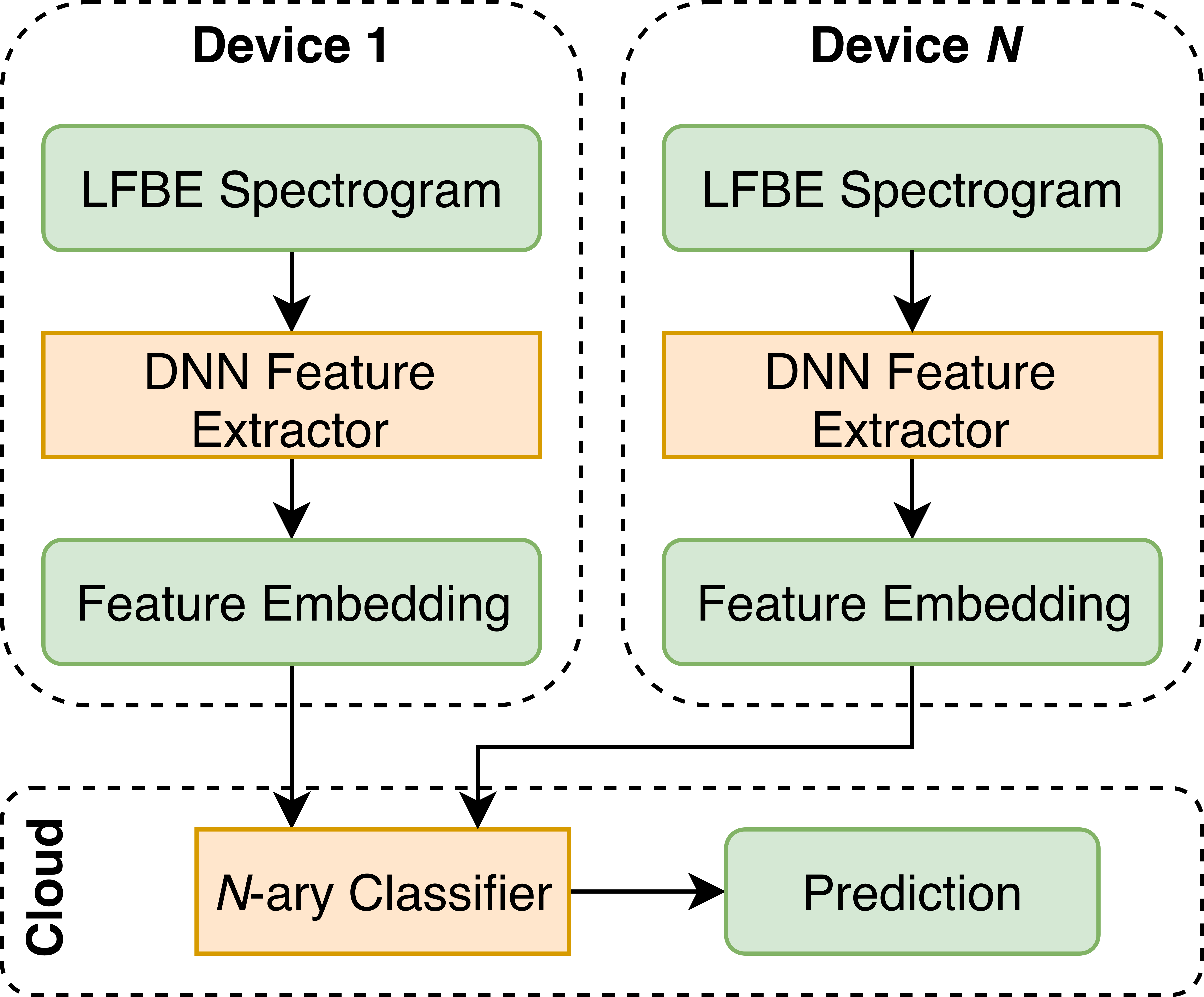}
  \vspace{-2mm}
  \caption{The end-to-end arbitration model architecture. We produce feature embeddings for each device, which are then fused in the cloud to produce a prediction. The entire system is trained end-to-end with simulated data.}
  \label{fig:e2e_arch}
\end{figure}

This yields a concrete technical problem: \emph{given short segments of audio (containing the wakeword) recorded from a set of $N$ (heterogeneous) microphone arrays, we want to determine which device is closest to the user}. Furthermore, this must be done without any a priori knowledge of the spatial relationships between the devices or the acoustic environment that they are in. We also desire the system to be able to operate on resource-constrained edge devices with minimal latency.

This problem can be seen as a variant of the source localization problem, which has been well-studied in the literature. When multiple microphones are present on a single device, time difference of arrival (TDOA) algorithms \cite{abel1987spherical,huang2001real,beck2008exact} can be used to estimate the range from the device to the speaker; however, these generally require physically large arrays to achieve the required accuracy for our problem. With the small footprints of consumer smart speakers, these methods become very sensitive to small errors in the TDOA estimates. TDOA between \emph{devices} may work in principal, but introduces other complexities such as clock and data synchronization between devices, and a solution that avoids these is preferable.

Another class of signal processing methods are \emph{energy-based methods} \cite{sheng2003energy,meesookho2007energy,meng2017energy}, which exploit the free-field relationship $E \propto 1/R^2$ between energy and range; when this holds, the closest device will have receive the maximum energy from the source. These methods face difficulties when applied to home smart speakers in reverberant rooms with high levels of noise and interference (such as TVs and air conditioners).

Most recently, deep learning methods have been applied to source localization; these attempt to directly regress the source location directly from features \cite{vesperini2016neural,ferguson2018sound,adavanne2018sound,chakrabarty2017broadband,quaternions}. 
This problem is also related to the channel selection problem \cite{wolf2014channel,cornell2021learning,cohen2014real}; however, the physically closest device may not actually provide the best channel, due to room effects and interference.

One limitation of applying the previous source localization algorithms to the device arbitration problem is that it is trying to solve a harder problem (estimating the range) than what we care about (finding the device with the smallest range). To overcome this limitation and directly solve the problem of interest, we propose an end-to-end neural network approach, which we illustrate in Figure \ref{fig:e2e_arch}. We train a common feature extractor that runs on each device, along with classification network that combines features from an arbitrary numbers of devices. We optimize the entire system to predict the optimal arbitration decision.

Due to the lack of large-scale arbitration datasets with known ground truth, we study this problem using a simulation of room and device acoustics. We build a pipeline for generating simulated rooms populated with devices, speakers, and noise sources, then computing both the device audio signals and spatial ground truth.

The rest of the paper is organized as follows: First, we detail our dataset simulation method for generating audio signals with spatial ground truth. Next, we discuss our end-to-end model. Third, we discuss our experiment setup, baseline, and results. Lastly, we conclude and discuss future work.
\section{Dataset Simulation}
\label{sec:sim}
\begin{figure}[t]
    \centering
    \includegraphics[width=0.7\linewidth]{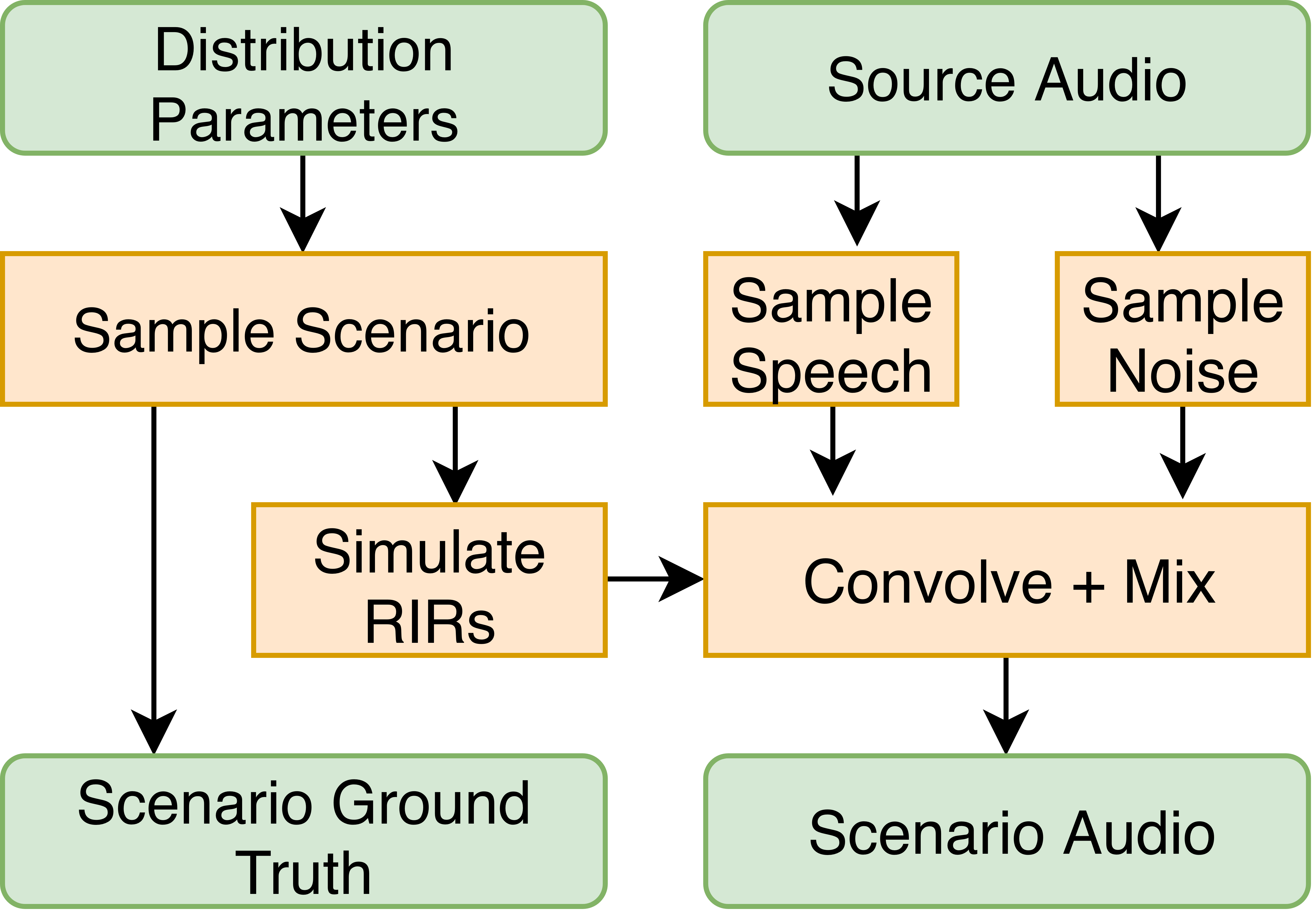}
    \vspace{-2mm}
    \caption{Pipeline for generating an arbitration scenario from a source audio dataset.}
    \label{fig:sim}
\end{figure}

Our dataset simulation pipeline is shown in Figure \ref{fig:sim}.
First, we sample \emph{arbitration scenarios} from a generative process. An arbitration scenario consists of a room with its physical and acoustic properties, the spatial locations of a number of devices placed in the room, and locations and dB SPL of both a human speaker and a number of noise sources. We will refer to these sampled parameters as \emph{arbitration metadata}.
We then use the metadata to simulate a set of room impulse responses (RIRs) between each sound source (speech/noise) and each device microphone, using a room simulator. Finally, we apply these RIRs to a \emph{source audio dataset} and mix.
This gives us the microphone signals for each device, along with the metadata that we need to compute our arbitration ground truth. The rest of this section details each of these sub-components.
\subsection{Generative process}
\begin{figure}[t]
  \centering
  \includegraphics[width=0.9\linewidth]{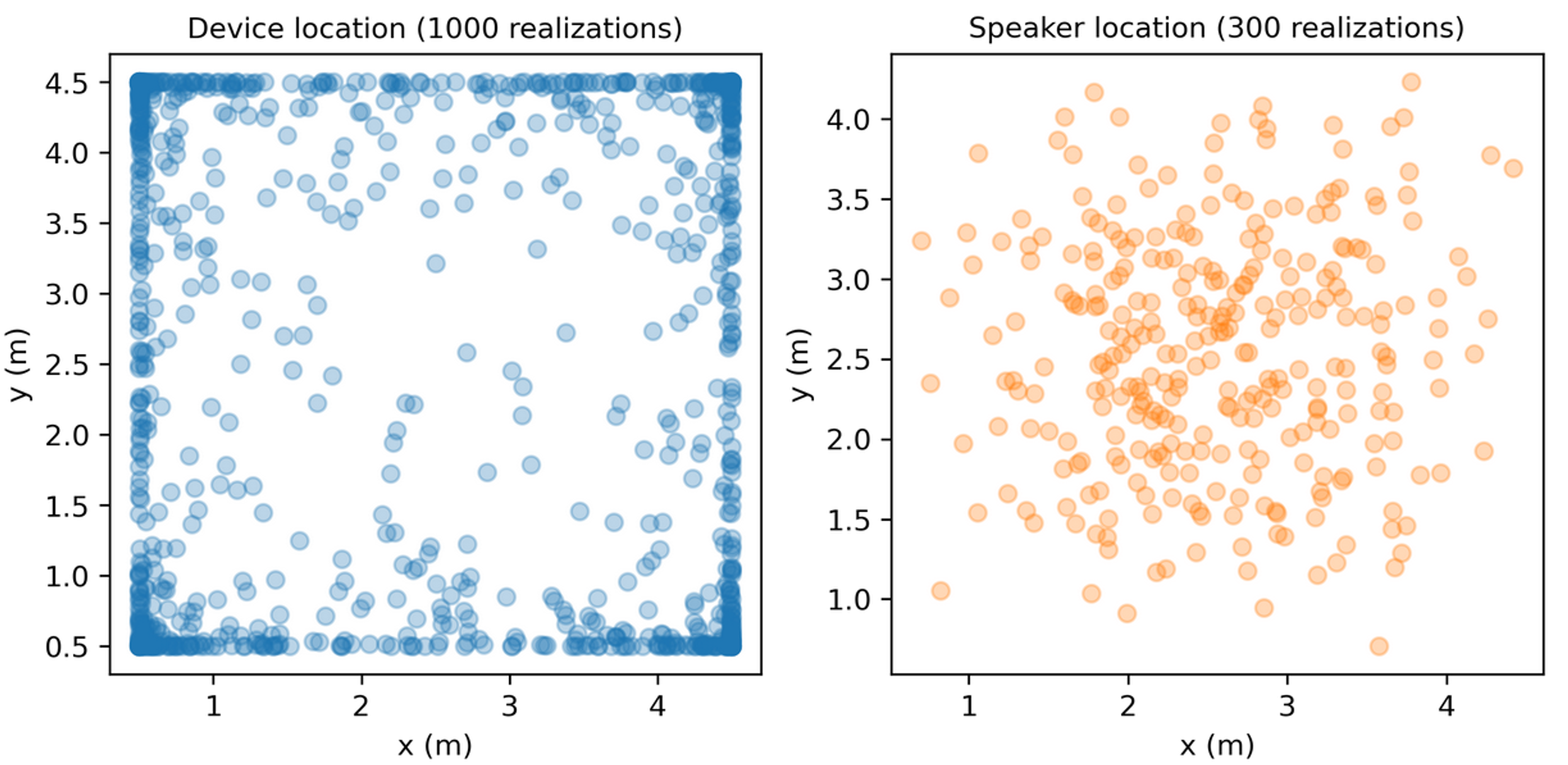}
    \vspace{-4mm}
  \caption{Samples of devices and speakers in a single room. We constrain the devices and speakers to be at least 10cm from a wall.}
  \label{fig:device_locs}
\end{figure}

We generate arbitration scenarios by sampling the scenario parameters from the distributions listed in Table \ref{tbl:gen_params}. The parameters for room size, reverberation time, and speech/noise levels are taken from \cite{kim2017generation}. First, we sample the room properties: its size and reverberation time. Next, we populate it with devices, speakers, and noise sources; Figure \ref{fig:device_locs} shows examples of how the devices and speakers are placed in a room. Finally, we apply rejection sampling to enforce a constraint that the minimum distance between the speaker and a device is at least 1 meter.
\begin{table}
    \centering
  \begin{tabular}{|c|c|c|}
    \hline
    \textbf{Parameter} &  \textbf{Distribution} \\ \hline\hline
    Room length/width (m) &  Uniform(3.0, 10.0) \\ \hline 
    Room height (m)&  Uniform(2.5, 6.0) \\ \hline 
    Reverberation time (s)&  Beta(2.5, 1.8) \\ \hline 
      Number of devices &  
      \makecell{Categorical: \\ P(2,3,4,5)=0.7,0.25,0.03,0.02} 
      \\ \hline 
    Device location & Beta(0.2, 0.2) \\ \hline 
    Speaker location & Beta(3.0, 3.0) \\ \hline 
    Number of noise sources & Poisson(2) \\ \hline
    Noise source location & Uniform(0, 1) \\ \hline
    Noise level (dB SPL)&  Uniform(25, 80) \\ \hline 
    Speech level (dB SPL) &  Uniform(45, 70) \\ \hline 
  \end{tabular}
  \caption{The distributions used to sample the arbitration scenario parameters. The device, speaker, and noise source location distributions are relative to the room size.}
  \label{tbl:gen_params}
\end{table}
\subsection{RIR simulation}
The room acoustics model that we use is a variant of the image source method \cite{allen1979image}, which we augment with high-fidelity COMSOL models of the farfield responses for different microphone arrays of interest. Specifically, 
our COMSOL models allow us to define a function $\phi_m(\hat{\vf r}, t)$ which gives the response of microphone $m$ to an impulse from farfield direction $\hat{\vf r}$.
The resulting RIR is given by a sum over image sources:
\begin{equation}
    h_m(t) \propto \sum_{k=0}^\infty \frac{\beta^{n_k}}{t_k^m} \phi_m(\hat{\vf r}_k^m, t - t_k^m)
  \label{eq:ism}
\end{equation}
where $\beta$ is the room absorption coefficient (estimated from the desired reverberation time \cite{zaheer2017deep}); $n_k$ is the order (number of reflections) of the $k$th image source; $t_k^m$ and $\hat{\vf r}_k^m$ are the time delay and DOA vector from microphone $m$ on the device to the image source. Using the per-microphone time delay is important for correctly capturing the wavefront curvature across the device. 
\subsection{Source audio}
We use two source audio datasets: Google SpeechCommands v2 (GSCv2) \cite{warden2018speech} and an internal source audio dataset (which we will refer to as the ``Alexa'' dataset).
For the GSCv2, we used 2377 utterances of the ``seven'' keyword and the provided background audio (11 minutes).
The internal audio consisted of 1000 recordings of the ``Alexa'' wakeword (male and female speakers, recorded in quiet but not perfectly anechoeic conditions) as the speech audio. For the background audio, we used internally-collected recordings of ambient background noise (air conditioners, running water, etc.) as well as 30 hours of TV show and movie audio from Amazon Prime Video. 
For both source audio datasets, the speech segments were partitioned into train/val/test splits, and the background audio segments were divided into non-overlapping subsegments before being assigned to train/val/test splits. All audio is sampled at 16kHz.

To generate the audio data for each device, we sample speech and interference segments from the appropriate source audio dataset. These are then convolved with the speech and background RIRs computed by the room simulator before being mixed together. Lastly, the audio for each device has an independent timing jitter applied (Gaussian with $\sigma=0.1$s) before being trimmed to a fixed length. The timing jitter mimics the lack of synchronization between devices.
\section{Method}
Our model consists of two components: a feature extractor, which runs on each device and produces a feature embedding, and a classifier, which runs in the cloud (or some other central hub) to compute the final prediction and make the arbitration decision (Figure \ref{fig:e2e_arch}).

As input data, we use a 2 second window of single-microphone audio; we assume that the wakeword speech is present somewhere in this window. Next, we convert this audio to log-filterbank energy (LFBE) features: we compute a spectrogram with a 25ms frame size and a 10ms frame skip, followed by a mel transform (with 64 mel bands) and a log transform. This results in a $201\times 64$ LFBE image that we feed to our network.

The feature extraction network consists of 5 convolution layers, followed by a single fully connected layer, with a total of 132k parameters. This produces a 128 dimensional embedding, which is passed to the $N$-ary classifier. The $N$-ary classifier is designed to map $N$ 128-dimensional embeddings to $N$ output probabilities. If we denote the feature embedding from device $j$ as $\vf z_j$, the network calculates the corresponding logit $\ell_j$ by: 
\begin{equation}
  \ell_j = g\left(\vf z_j, \sum_i \vf z_i\right)
\end{equation}
where $g$ is a two-layer fully connected neural network. The final network output is given by applying a softmax function to the vector of logits.
This approach to fusing the embeddings of the other devices to make a permutation-equivariant scoring function is maximally generic (see Theorem 2 in \cite{zaheer2017deep}).

The feature extraction and $N$-ary classifier networks are simultaneously trained end-to-end using Adam and the standard cross-entropy loss, where ground truth is extracted from the scenario metadata.
\begin{table}[t]
    \centering
    \begin{tabular}{|c|c|c|}
        \hline
        \textbf{Train Dataset} & \textbf{Test Dataset} & \textbf{Rel. error rate} \\ \hline
            \multirow{2}{*}{Alexa} & Alexa & 52.8\% \\ 
             & GSCv2 & 60.1\% \\  \hline\hline
            \multirow{2}{*}{GCSv2} & GCSv2 & 50.6\% \\ 
             & Alexa & 63.7\% \\ \hline
    \end{tabular}
    \caption{Arbitration relative error rate of the end-to-end DNN over the baseline. We test each model on both test sets to get a sense of how sensitive the results are to the specific keywords.}
    \label{tbl:main_acc}
\end{table}
\section{Experiments}
\subsection{Baseline}
\label{sec:baseline}
We use a simple energy-based method as a baseline. Denoting the microphone signal (assumed to contain the wakeword) from device $k$ as $\vf x_k$, our decision rule becomes:
\begin{equation}
    k_\text{closest} = \argmax_{k} \|\vf h_\text{bandpass} * \vf x_k\|^2
\end{equation}
where $\vf h_\text{bandpass}$ is a bandpass filter that passes 1500Hz through 6500Hz. These parameters were chosen by optimizing accuracy over a validation set.
\subsection{Experimental procedure}
For each source audio dataset (GSCv2 and our Alexa dataset), we generate train/val/test sets of 50k/2k/10k scenarios, respectively; we used the same scenarios for each source audio dataset. We also generate a separate dataset (from the same distribution) of 10k \emph{noise free} scenarios, which we used in an ablation to evaluate our baseline in ideal conditions. We train our end-to-end model on the training set for 30 epochs and take the model with the best validation accuracy. We then compare the results against the baseline (Section \ref{sec:baseline}) using 3 metrics: overall accuracy (probability of correctly identifiying the closest device), $\Delta$-accuracy, and $\epsilon$-accuracy. We define these last two metrics next.

For a given scenario, let $d_1$ be the distance of the closest device to the user and $d_2$ be the distance of the \emph{second} closest device to the user. Then we define $\Delta$-accuracy to be the accuracy when the difference $d_2 - d_1$ is equal to $\Delta$ (where $\Delta \ge 0$ is a given parameter). This captures the idea that low-$\Delta$ scenarios are both harder (more ambiguous) and less impactful to the user's experience - it is worse to fail in cases where $\Delta$ is large than when it is small.

Similarly, let $d$ be the distance of the \emph{chosen} device (e.g., the argmax of the classifier output) to the user. We define $\epsilon$-accuracy, where $\epsilon \ge 0$, as:
\begin{equation}
    \epsilon\text{-accuracy} \overset{\text{def}}{=} P(d - d_1 < \epsilon)
\end{equation}
In other words, this is the probability of choosing a device that is $\epsilon$-close to the optimal device. The case of $\epsilon=0$ coincides with the normal accuracy metric. This metric is most useful with scenarios of 3 or more devices; for example, if the distance from each device to the user is 2.0m, 2.2m, and 6.0m, $\epsilon$-accuracy will capture the idea that choosing the second device is better than choosing the third.
\section{Results}
\begin{figure}[t!]
    \centering
    \includegraphics[width=0.95\linewidth]{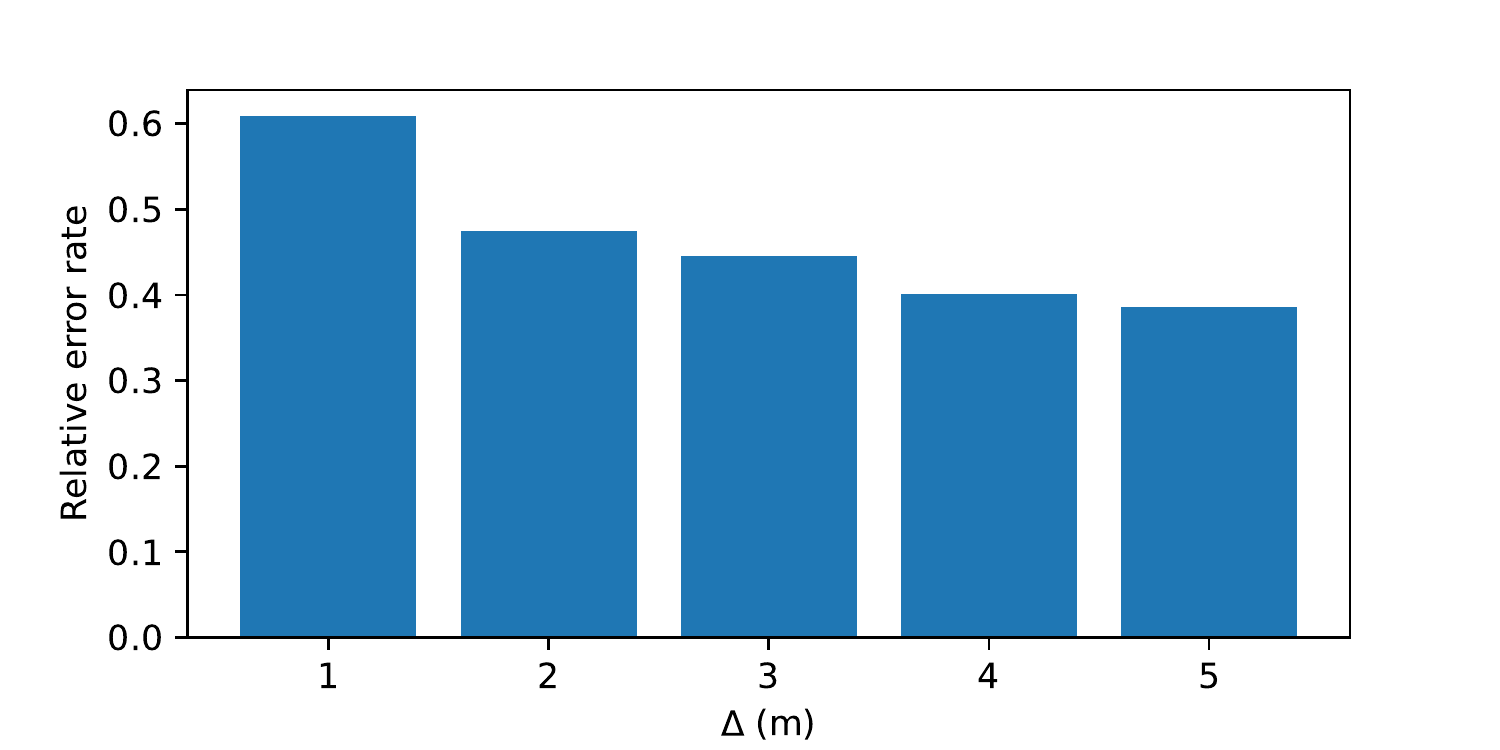}
    \vspace{-2mm}
    \caption{Relative error rate as a function of $\Delta$, binned in 1m increments. Our model uniformly outperforms the baseline for all values of $\Delta$.}
    \label{fig:delta_acc}
\end{figure}

\begin{figure}
    \centering
    \includegraphics[width=0.95\linewidth]{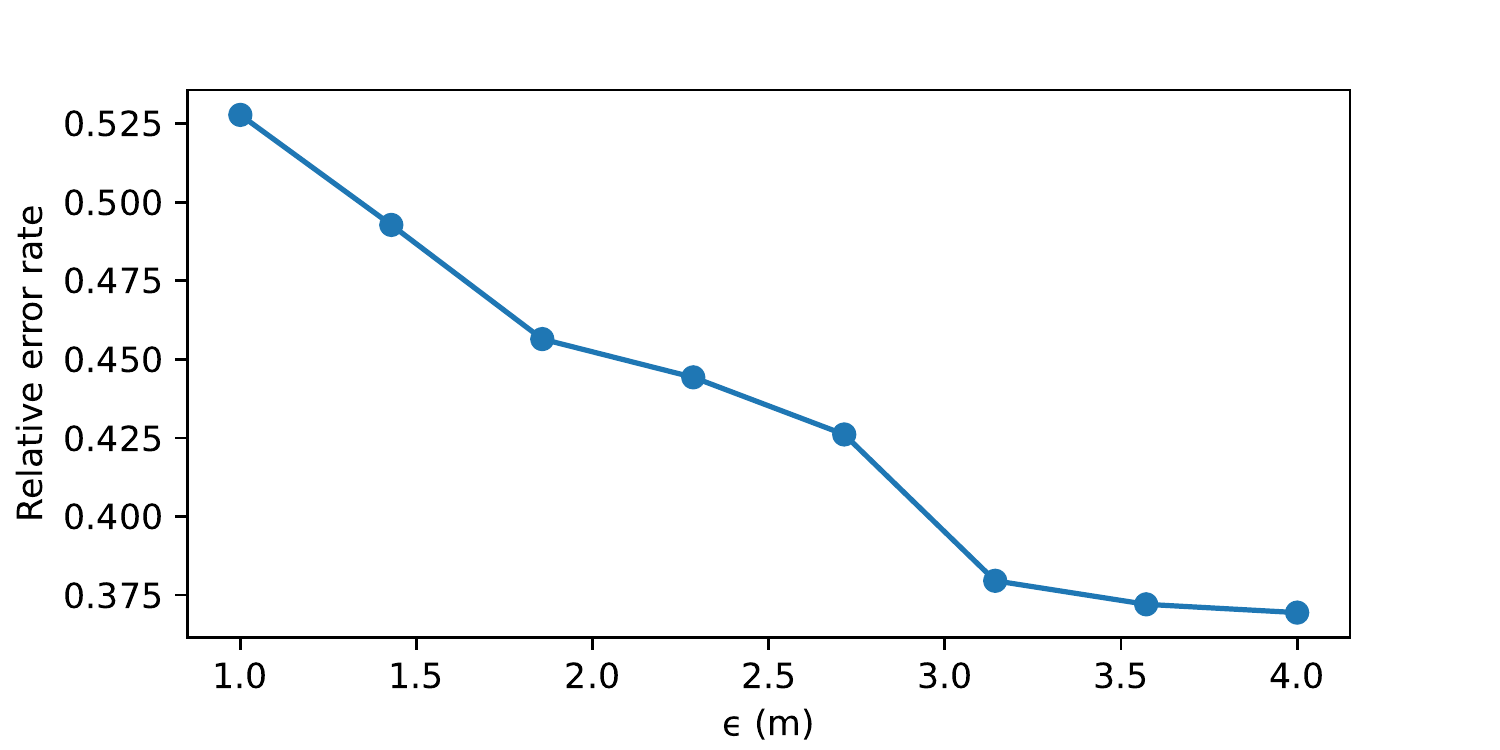}
    \vspace{-2mm}
    \caption{Relative error rate as a function of $\epsilon$. Our model uniformly outperforms the baseline for all values of $\epsilon$.}
    \label{fig:eps_acc}
\end{figure}

We present our results in terms of \emph{relative error rate} over the baseline. Given the accuracies $\text{acc}_\text{bl}, \text{acc}_\text{dnn}$ of the baseline model and the DNN model, respectively, the relative error rate is computed via:
\begin{equation}
    \text{err}_\text{rel} = \frac{1 - \text{acc}_\text{dnn}}{1 - \text{acc}_\text{bl}}
\end{equation}
Our main results on are shown in Table \ref{tbl:main_acc}. Our end-to-end method substantially outperforms the baseline, even though the baseline method was tested in a noise free environment, and reduces classification error by 49\% and 47\% for the GSCv2 and Alexa datasets, respectively. We also test on the other dataset to get a sense of out-of-domain generalization (e.g., a completely different wakeword) and still see substantial improvement over baseline, although less than the in-domain case. Our model also exhibits relative performance improvements on both our $\Delta$-accuracy (Figure \ref{fig:delta_acc}) and $\epsilon$-accuracy (Figure \ref{fig:eps_acc}) metrics vs. the baseline for all values of $\Delta$ and $\epsilon$. 



We ran three additional ablations: First, we tried a non-end-to-end version of the model where we independently predicted the range to the speaker on each device, resulting in worse performance. This is expected, since predicting the absolute range should be harder than determining the closest device. Second, we experimented with using data from multiple microphones, with both LFBE features as well as complex STFT and cross-correlations between channels. None of these resulted in better performance than single-mic LFBEs. We conjecture that while phase information is critical for DOA estimation, it does not carry much information about range, at least for small arrays. Third, we looked at baseline performance on a noise-free dataset. We found only very small differences in accuracy ($<$ 1\%), suggesting that room reverberation is the dominant source of baseline error.
\section{Conclusion}
In this paper, we have presented an end-to-end deep learning system for device arbitration among smart speakers. We use simulation to generate large datasets with spatial ground truth, which we use to generate training and evaluation data for our model. Our model shows substantial performance improvements over our signal-processing based baseline.

We have several directions for future work. One is to collect data in real-world acoustic environments (where spatial ground truth is known) to validate and possibly train our models. We are also exploring improving our room simulator to support multi-room environments as well as model speaker directivity \cite{monson2012horizontal}. Lastly, we are exploring self-supervised learning to fuse representations learned from real data with the high-fidelity ground truth that our simulations provide, with the goal of improving performance and generalization.
\vfill\newpage
\nocite{*}
\bibliographystyle{IEEEbib}
\bibliography{refs}

\end{document}